\documentclass[%
 reprint,
superscriptaddress,
 amsmath,amssymb,
]{revtex4-2}

\usepackage{graphicx}
\usepackage{dcolumn}
\usepackage{bm}
\usepackage{bbm}
\usepackage[caption=false]{subfig}
\usepackage{float}
\usepackage{hyperref}
\usepackage{multirow}
\usepackage{algorithm}
\usepackage{algpseudocode}

\usepackage{amsthm}

\begin{document}

\preprint{APS/123-QED}

\title{Tight Secure Key Rates for CV-QKD with 8PSK Modulation}
\author{Florian Kanitschar}
\affiliation{AIT  Austrian  Institute  of  Technology,  Center  for  Digital  Safety\&Security,  Giefinggasse  4,  1210  Vienna, Austria}
 \affiliation{TU Wien, Faculty of Physics, Wiedner Hauptstraße 8, 1040 Vienna, Austria}
 %\email{florian.kanitschar@ait.ac.at}
 \author{Christoph Pacher}
  \affiliation{AIT  Austrian  Institute  of  Technology,  Center  for  Digital  Safety\&Security,  Giefinggasse  4,  1210  Vienna, Austria}
 %\email{christoph.pacher@ait.ac.at}

\date{\today}

\begin{abstract}
We use a recent numerical security proof approach to calculate tight secure key rates in the presence of collective attacks for a continuous-variable (CV) eight-state phase-shift keying (8PSK) protocol with heterodyne detection in the asymptotic limit. The results are compared to achievable secure key rates of a QPSK protocol obtained with the same security proof technique. Furthermore, we investigate the influence of radial postselection on the secure key rate and examine a recently suggested strategy to reduce the computational requirements on the error-correction phase for four-state phase-shift-keying protocols. Based on our investigations, we suggest different strategies to reduce the raw key of 8PSK protocols significantly, only on the cost of a slightly lower secure key rate. This can be used to lower the computational effort of the error-correction phase, a known bottleneck in many practical implementations, considerably.

\end{abstract}

\maketitle

%\tableofcontents

%--------------------------------------------------------
%                    Introduction
%--------------------------------------------------------

\section{\label{sec:Intro}Introduction}
Quantum key distribution, in short QKD, aims to provide a random bit sequence called secret key for two remote parties commonly named Alice and Bob \cite{Pirandola_2020, Diamanti_2015, Scarani_2009}. It is based only on the laws of quantum physics and can be proven  information theoretically secure, assuming an eavesdropper called Eve who can read out and manipulate the quantum channel connecting Alice and Bob and is assumed to have access to unlimited computational power.\\

There are two main branches for QKD-protocols, namely discrete-variable (DV) QKD and continuous-variable QKD. Bennett and Brassard's seminal protocol BB84 \cite{Bennett_Brassard_1984} relies on discretly polarized photons and is a member of the family of DV-QKD protocols. In CV QKD protocols, the information is carried by continuous quantities like the field quadratures $\hat{q}$ and $\hat{p}$. They can be operated using commercially available hardware, commonly used in modern telecommunication-infrastructure, but it remains a theoretical challenge to prove general security, i.e., to calculate or lower-bound the secure key rate. CV QKD protocols are divided into two sub-branches, depending on the modulation method. The most common modulations are Gaussian, where Alice chooses the amplitude of the sent coherent states according to a Gaussian distribution, and discrete modulation, where Alice chooses some small number of coherent state amplitudes. Complete security proofs for Gaussian modulated CV-QKD protocols for a finite sized key against general attacks are available \cite{Leverrier_2017}. For discretely modulated CV-QKD protocols, there are also security proofs against collective attacks in the asymptotic limit available \cite{Bradler_2018, Lin_2019, Ghorai_2019}. Recent attempts \cite{George_2021, Matsuura_2021, Bunandar_2020} tackle the open problem of secure key rates in the finite-size regime. \\

In the present work, we use the security proof approach from \cite{Winick_2018, Lin_2019} and examine secure key rates for phase-shift keying protocols with 8 signal states (8PSK) both with and without postselection for different practically relevant values of excess-noise and reconciliation efficiency. We extend examinations for four-state protocols from our previous work \cite{Kanitschar_2021} to eight-state protocols, using generalised analytical expressions. Furthermore, we point out how postprocessing can not only be used to increase the secure key rate, but also to reduce the amount of raw key significantly while decreasing the secure key rate only moderately. This reduces the amount of data that has to be error-corrected considerably, hence is beneficial for practical implementations, as it lowers the computational requirements for the error-correction phase, which is a well-known bottleneck in many experiments.\\

The remainder of this work is structured as follows. In Section \ref{sec:protocol}, we introduce the examined eight-state protocol. This is followed by a revision of the used numerical security proof method in Section \ref{sec:proof_method}. In Section \ref{sec:remarks}, we give details about the implementation and list the used numerical parameters. In Section \ref{sec:Validation}, we compare the results of our implementation for the special case of no excess noise with analytical results. In Section \ref{sec:results}, we compare the key rates for the eight-state protocol with results for a four-state phase-shift keying protocol from an earlier work \cite{Kanitschar_2021} and examine the influence of radial postselection on the secure key rate. In particular, we point out how postselection can be used to reduce the raw key significantly without decreasing the secure key rate considerably. This is followed by a brief discussion of our findings in Section \ref{sec:Conclusion}.
%--------------------------------------------------------
%             Introduction of the Protocol
%--------------------------------------------------------

\section{Introduction of the protocol}\label{sec:protocol}
We examine an eight-state phase-shift keying protocol which is the natural generalisation of the four state-protocol described in \cite{Lin_2019}. The communicating parties perform $N\in \mathbb{N}$ rounds of key generation, following the instructions listed below. By $n \leq N$ we numerate the rounds of key generation.

\begin{enumerate}
\item[1) ] Alice chooses some arbitrary but fixed coherent state amplitude $|\alpha| > 0$ (her choice may depend on parameters like her distance to Bob, or the channel-noise) and prepares one of out of eight coherent states $\left| \Psi_k \right\rangle = \left| |\alpha| e^{i k\frac{\pi}{4} } \right\rangle$ corresponding to the symbol $x_n = k$, $k \in \{0,...,7\}$, with probability $p_k$ (see Figure \ref{fig:sketch_phasespace}). For symmetry reasons, we choose the uniform distribution, so $\forall k \in\{0,...,7\}: ~ p_k =  \frac{1}{8}$. Then, she sends this state to Bob via the quantum channel.

\begin{figure}
\subfloat[ \label{fig:sketch_phasespace}Sketch of the prepared states in phase-space]{
\includegraphics[width=0.48\textwidth]{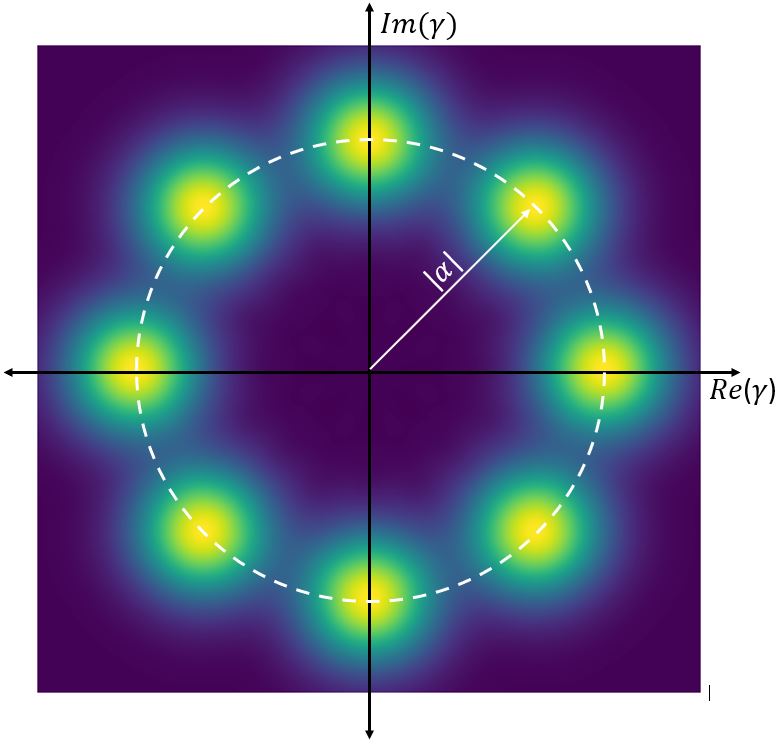}}\\
\subfloat[\label{fig:key_map} Sketch of the 8PSK key map including postselection areas (blue). We denote by $\Delta_r$ the radial postselection parameter. So, Bob's measurement results that lie in one of the blue-shaded areas are postselected. Formally, they are assigned to the symbol $\perp$, while the remaining results are assigned to the number written in the corresponding region.]{
\includegraphics[width=0.48\textwidth]{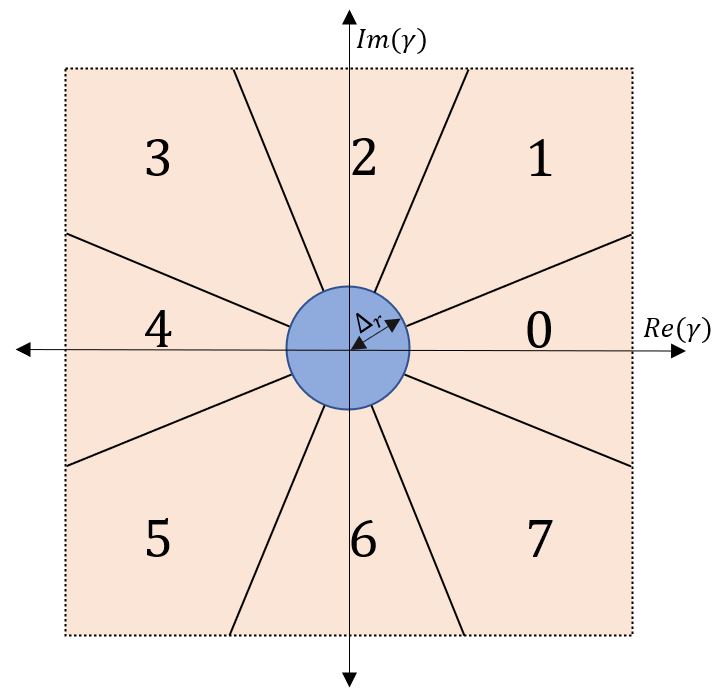}}
\caption{Sketches for the 8PSK protocol}
\end{figure}

\item[2) ] Once Bob receives the quantum state, he performs a heterodyne measurement. His measurement outcome is some complex number $y_n \in \mathbb{C}$.

\item[3) ] For parameter estimation, Alice and Bob agree on some randomly chosen subset $\mathcal{I}_{\text{Test}} \subset \{n \in \mathbb{N} ~: ~ n \leq N\}$ and announce the corresponding symbols $x_l$ and results $y_l$ for $l \in \mathcal{I}_{\text{Test}}$ to determine Eve's potential knowledge about the key. Consequently, they use the remaining rounds $\mathcal{I}_{\text{key}} := \{n \in \mathbb{N} ~:~n\leq N\}\setminus \mathcal{I}_{\text{Test}}$ for key generation. To ease the notation, we assume (without loss of generality) that the first $m := |\mathcal{I}_{\text{key}}|$ rounds can be used for key generation, such that, finally, Alice holds a key string $\mathbf{X} := (x_1,...,x_m)$.

\item[4) ] Bob determines his key string $\mathbf{Z} = (z_j)_{j \in \mathcal{I}_{\text{key}}}$ using a reverse reconciliation key map (see Figure \ref{fig:key_map}). Therefore, the phase-space is partitioned into eight regions, where each of them is associated with a number from $0$ to $7$, and a postselection area that is associated with the symbol $\perp$. Then, for $k \in\{0,...,7\}$, Bob obtains his key string as follows
\begin{align}
z_j = \left\{
\begin{array}{ll}
k & \text{if } \arg(y_j) \in \left[\frac{(2k-1)\pi}{8}, \frac{(2k+1)\pi}{8}\right) \land |y_j| \geq \Delta_r, \\
\perp &  \textrm{otherwise.} \\
\end{array}
\right. 
\end{align}
Consequently, the corresponding postselection areas can be described by the following sets
\begin{equation}\label{eq:regions}
\begin{aligned}
    &A_k^{\text{r}} := \\
    &\left\{ z \in \mathbb{C}: \text{arg}(z) \in \left[ \frac{(2k-1)\pi}{8}, \frac{(2k+1)\pi}{8} \right) \land |z|\geq \Delta_r \right\}.
\end{aligned}
\end{equation}

\item[5) ] In the end, the communicating parties perform classical error correction and privacy amplification algorithms.  
\end{enumerate}

%--------------------------------------------------------
%             Security proof method
%--------------------------------------------------------
\section{Description of the security proof method}\label{sec:proof_method}
\subsection{Formulation of the key rate finding problem}
We use the numerical security proof approach from \cite{Winick_2018, Lin_2019}, where the expression for the secure key rate in the asymptotic limit follows from a reformulation of the Devetak-Winter formula \cite{Devetak_Winter_2006},
\begin{equation}
R^{\infty} = \min_{\rho_{AB} \in \mathcal{S}} D\left(\mathcal{G}(\rho_{AB}) || \mathcal{Z}(\mathcal{G}(\rho_{AB}))\right) - p_{\text{pass}} \delta_{EC}.    
\end{equation}
Here $D(\rho||\sigma) := \text{Tr}\left[\rho \left(\log_2(\rho)-\log_2(\sigma)\right)\right]$ is the quantum relative entropy, which is a distinguishability-measure between the states $\rho$ and $\sigma$, $\mathcal{G}$ is a completely-positive trace non-increasing map and $\mathcal{Z}$ is a pinching quantum channel. Both maps will be specified in what follows. Finally, $\delta_{EC}$ is the information leakage per signal in the error-correction phase and $p_{\text{pass}}$ is the probability that a signal passes the postselection phase. The minimisation is carried out over the set $\mathcal{S}$, which is a subset of the set of all positive semi-definite operators $\mathcal{H}_+$ and defined by a number of linear constraints.\\
The postprocessing map $\mathcal{G}$ is defined as $\mathcal{G}(\sigma) = K \sigma K^{\dagger}$, where $K$ is a Kraus operator, defined by
\begin{equation}
    K := \sum_{z=0}^7 |z\rangle_R \otimes \mathbbm{1}_A \otimes \left(\sqrt{R_z} \right)_B,
\end{equation}
and where $\left(R_z\right)_{z \in\{0,...,7\}}$ are the so-called region operators. If $E_y$ is the POVM of the underlying heterodyne measurement, they read 
\begin{equation}\label{eq:RegionOpDef}
    R_z := \int_{\mathcal{A}_z^{\text{r}}} E_y \,d^2y,
\end{equation}
where $\mathcal{A}_z^{\text{r}}$ is a set, describing the region in the phase-space corresponding to the symbol $z$, as defined eq. (\ref{eq:regions}).  
The pinching channel $\mathcal{Z}$ is given by
\begin{equation}
\mathcal{Z}(\sigma) := \sum_{j=0}^{7} \left(|j\rangle\langle j|_R \otimes \mathbbm{1}_{AB} \right) \sigma \left(|j\rangle\langle j|_R \otimes \mathbbm{1}_{AB} \right).
\end{equation}
It remains to define $p_{\text{pass}}$ and $\delta_{EC}$ in order to fully specify the objective function. The region operators $R_z$ describe Bob's measurement associated with a certain symbol $z$. Therefore, we express the probability that Bob obtains the symbol $z=k$ conditioned that Alice has prepared the state $x=l$ by
\begin{equation}\label{eq:pCond}
P(z=k |x=l ) = \text{Tr}\left[\rho_B^l R_k\right],
\end{equation}
where 
\begin{equation}
    \rho_B^l = \frac{1}{p_l} \text{Tr}_A\left[ \rho_{AB} \left(|l\rangle\langle l|_A \otimes \mathbbm{1}_B \right) \right]
\end{equation}
is the state Bob receives, conditioned that Alice prepared the state corresponding to the symbol $x=l$. Assuming that we can perform information reconciliation at the Slepian-Wolf limit \cite{Slepian_Wolf_1973} we would have $\delta_{EC} = H(\mathbf{Z} |\mathbf{X}) = H(\mathbf{Z}) - I(\mathbf{X}:\mathbf{Z})$, where $H(\mathbf{Z})$ is the von-Neumann entropy of the bit-string $\mathbf{Z}$, $H(\mathbf{Z}|\mathbf{X})$ is the conditioned von-Neumann entropy between the bit-strings $\mathbf{X}$ and $\mathbf{Z}$, and $I(\mathbf{X}:\mathbf{Z})$ denotes the mutual information. Unfortunately, this is not possible in practical implementations, where the information reconciliation efficiency $\beta < 1$. Therefore, we replace the mutual information in the expression above by $\beta I(\mathbf{X}:\mathbf{Z})$ and insert back entropies again. This leads to 
\begin{equation}
\begin{aligned}
\delta_{EC} &= H(\mathbf{Z}) - \beta \left(H(\mathbf{Z}) - H(\mathbf{Z}|\mathbf{X})\right)\\
&= (1-\beta) H(\mathbf{Z}) + \beta H(\mathbf{Z}|\mathbf{X}),
\end{aligned}
\end{equation}
where the occurring entropies can be calculated using Eq. (\ref{eq:pCond}) and basic probability theory.\\
Denoting $p_l$ the probability that Alice prepares the state associated with $l$ (which we assume to be $\frac{1}{8}$ for all $l$ for the present protocol), we obtain the probability to pass the postselection phase,
\begin{equation}
    p_{\text{pass}} = \sum_{l=0}^{7} \sum_{k=0}^{7} p_l P(z=k|x=l),
\end{equation}
which, again, can be calculated using eq. (\ref{eq:pCond}) and basic probability theory.\\

It remains to specify the feasible set of the optimisation problem $\mathcal{S}$. Therefore, we quickly revisit the state-generation and transmission process. First, Alice prepares one of the states $\{|\psi_x\rangle\}_{x \in \{0,...,7\}}$ with probability $p_x$ and sends them to Bob via the quantum channel. Following the source-replacement scheme \cite{Curty_2004, Ferenzci_2012}, equivalently, one can consider the corresponding entanglement-based scheme, where Alice prepares the bipartite state $|\Psi\rangle_{AA'} = \sum_{x} \sqrt{p_x} |x\rangle_A |\psi_x\rangle_{A'}$. She decides to keep one share, denoted by $A$, and sends the one in register $A'$ to Bob. 
The quantum channel connecting Alice's and Bob's labs is described by a completely positive trace preserving map $\mathcal{E}_{A'\rightarrow B}$. Therefore, after the transmission, they hold the state $\rho_{AB} = \left(\mathbbm{1}_A \otimes \mathcal{E}_{A'\rightarrow B} \right)\left(|\Psi\rangle\langle \Psi|_{AA'} \right)$. Then, Bob performs heterodyne measurement on his share, hence determines the first- and second-moments of $\hat{q}$ and $\hat{p}$, or, equivalently, the first-moments of $\hat{q}$ and $\hat{p}$ and the following quantities that can be derived from the second moments, $\hat{n} = \frac{1}{2} \left(\hat{q}^2 + \hat{p}^2 -1 \right)$ and $\hat{d} = \hat{q}^2 - \hat{p}^2$.\\
Furthermore, Alice's state does not leave her lab, hence is inaccessible for Eve. Mathematically, this fact can be expressed by $\text{Tr}_B\left[ \rho_{AB} \right] = \sum_{x,y = 0}^{7} \sqrt{p_x p_y} \langle \psi_y|\psi_x\rangle~ |x\rangle \langle y|_A$, which is a matrix-valued constraint.\\
Finally, this considerations lead us to the following semi-definite program \cite{Lin_2019}
\begin{equation} \label{eq:SDP}
\begin{aligned}
    \text{minimise } &D(\mathcal{G}(\rho_{AB}) || \mathcal{G}(\mathcal{Z}(\rho_{AB}))  )\\
    \text{subject to: } &\\
    &\text{Tr}\left[ \rho_{AB} \left( |x\rangle\langle x|_A \otimes \hat{q} \right) \right] = p_x \langle \hat{q} \rangle_x\\
    &\text{Tr}\left[ \rho_{AB} \left( |x\rangle\langle x|_A \otimes \hat{p} \right) \right] = p_x \langle \hat{p} \rangle_x\\
    &\text{Tr}\left[ \rho_{AB} \left( |x\rangle\langle x|_A \otimes \hat{n} \right) \right] = p_x \langle \hat{n} \rangle_x\\
    &\text{Tr}\left[ \rho_{AB} \left( |x\rangle\langle x|_A \otimes \hat{d} \right) \right] = p_x \langle \hat{d} \rangle_x\\
    &\text{Tr}_B\left[ \rho_{AB} \right] = \sum_{i,j=0}^{7} \sqrt{p_i p_j} \langle \psi_j | \psi_i\rangle ~|i\rangle\langle j|_A\\
    & \rho_{AB} \geq 0,
\end{aligned}
\end{equation}
where $x \in \{0,...,7\}$.  Like in \cite{Kanitschar_2021}, we do not add the constraint $\text{Tr}\left[ \rho_{AB}\right] = 1$, requiring that $\rho$ has trace equal to $1$ explicitly as it is a density matrix explicitly, because we transform the matrix-valued constraint by quantum state-tomography (see, e.g., \cite{Altepeter_2005}) into a set of $64$ scalar-valued constraints. Together with the constraints corresponding to Bob's measurements, this gives us a set of constraints that is sufficient to linear-combine the trace-equal-to-one condition with sufficient numerical precision. For numerical reasons, it turned out to be beneficial to avoid constraints that are almost linearly-dependent. Furthermore, we remark that $p_{\text{pass}}$ is contained implicitly in the first term of the target function. A more detailed explanation is given in \cite{Lin_2019}.\\

For a phase-invariant Gaussian channel with transmittance $\eta$ and excess-noise $\xi$, the right-hand sides of the constraints due to Bob's measurements read
\begin{align}
    \langle \hat{q} \rangle_x &= \sqrt{2\eta}~ \Re(\alpha_x), \\
    \langle \hat{p} \rangle_x &= \sqrt{2\eta}~ \Im(\alpha_x), \\
    \langle \hat{n} \rangle_x &= \eta |\alpha_x|^2 + \frac{\eta \xi}{2},\\
    \langle \hat{d} \rangle_x &= \eta \left(\alpha_x^2 + (\alpha_x^*)^2 \right)
\end{align}
for $x \in \{0,...,7\}$ and $\alpha_x$ is a complex number associated with the coherent state Alice prepares. These expectation values, of course, are the same as for four-state protocols, hence can be found in \cite{Lin_2019}.\\
Denoting the Hermitian operator associated with the $i$-th constraint by $\Gamma_i$ and the corresponding right hand-sides of the $i$-th constraint by $\gamma_i$, we obtain the feasible set $\mathcal{S}$, a subset of the density operators $\mathcal{D}(\mathcal{H}_{AB})$, where $\mathcal{H}_{AB} = \mathcal{H}_A \otimes \mathcal{H}_B$. It can be desribed as follows
\begin{equation}
\mathcal{S} := \left\{ \rho_{AB} \in \mathcal{D}(\mathcal{H}_{AB}) ~|~ \forall i \in I: \text{Tr}\left[ \Gamma_i \rho_{AB} \right] = \gamma_i  \right\},
\end{equation}
where $I$ the index set of the constraints.\\

\subsection{Solving the occurring optimisation problem}
The present minimisation problem is a semi-definite program with non-linear objective function, living in an infinite-dimensional vector space. Therefore, Bob's Hilbert space $\mathcal{H}_B= \{|n\rangle ~ : ~ n \in \mathbb{N}\}$ is approximated by the subset, spanned by the first $N_c$ Fock-states, $\mathcal{H}_B^{N_c} := \text{span}\{|n\rangle ~ : ~ 0 \leq n \leq N_c\}$. The number $N_c \in \mathbb{N}$ is called the cutoff number and one can assume that the error we make is negligible \cite{Lin_2019}. This can be validated by monitoring the change in the key rate when increasing the cutoff number $N_c$. For all plots in the present thesis, we chose $N_c$ such that an increase did not lead to significant chances in the secret key rate. Recently, the cutoff-assumption could be removed \cite{Upadhyaya_2021}. \\

As we minimise numerically, we cannot expect to reach the minimum perfectly, which would be required for a valid security proof. Therefore, the problem is tackled by a two-step process \cite{Winick_2018}, where in the first step the problem is solved approximately and in a second step the obtained result (which is an upper bound on the secure key rate) is converted into a lower bound using a sequence of theorems, taking numerical imprecisions into account. Since the target function $f$ of the present optimisation problem is non-linear, we approximate it to first order and solve the linearised minimisation problem iteratively, using a modified Frank-Wolfe algorithm \cite{Frank_Wolfe_1956}, as suggested by \cite{Winick_2018}. 
\begin{algorithm}[H]
  \caption{Modified Frank-Wolfe for step 1}
  \label{ALG:FW}
   \begin{algorithmic}[1]
   \State Choose $\epsilon_{FW} > 0$, $\rho_0 \in \mathcal{S}$ and set $k=0$
   \State Find $\Delta \rho := \text{arg min}_{\Delta \rho} \text{Tr}\left[\left(\Delta \rho\right)^{\top} \nabla f(\rho_k)\right]$ subject to $\rho_k+ \Delta \rho \in \mathcal{S}$
   \State STOP if $\text{Tr}\left[\left(\Delta \rho\right)^{\top} \nabla f(\rho_k)\right] < \epsilon_{FW}$
   \State Find $\lambda \in (0,1) $ that minimises $f(\rho_k + \lambda \Delta \rho)$
   \State $\rho_{k+1} := \rho_{k} + \lambda \Delta \rho$, $k \leftarrow k+1$, proceed with 2.
   \end{algorithmic}
\end{algorithm}
Linearisation involves the evaluation of the gradient of the objective function, which is not guaranteed to exist on the whole domain of optimisation (for example, if the map $\mathcal{G}$ does not have full rank). Therefore, one may introduce a perturbed map
\begin{equation}
    \mathcal{G}_{\tilde{\epsilon}}(\rho) := \mathcal{D}_{\tilde{\epsilon}}(\mathcal{G}(\rho)),
\end{equation}
where $0 < \tilde{\epsilon} < 1$ and 
\begin{equation}
\mathcal{D}_{\tilde{\epsilon}}(\rho) := (1-\tilde{\epsilon}) \rho + \tilde{\epsilon} \frac{1}{\text{dim}(\mathcal{G}(\rho))}\mathbbm{1}_{N_c},
\end{equation}
resulting into a differentiable perturbed target function $f_{\tilde{\epsilon}}(\rho) := D \left( \mathcal{G}_{\tilde{\epsilon}}(\rho) || \mathcal{Z}( \mathcal{G}_{\tilde{\epsilon}}(\rho)  ) \right)$ \cite{Winick_2018}. To ease the notation, in what follows, we replace every $\mathcal{G}$ by $\mathcal{G}_{\tilde{\epsilon}}$ and $f$ by $f_{\tilde{\epsilon}}$ but do not state the subindex explicitly.\\
The second step converts the result from step 1, which is only an upper bound on the secure key rate, into a lower bound, taking differences between exact constraints and their computer representation into account. Denoting the computer representation of the Hermitian operators $\Gamma_i$ by $\tilde{\Gamma}_i$ and the computer representation of the right hand-sides by $\tilde{\gamma}_i$. The following theorem, given in \cite{Winick_2018}, states how one can obtain a lower bound the secure key rate, given that the constraints are satisfied up to some small number $\epsilon' \in \mathbb{R}$, $\forall i \in I: ~\left| \text{Tr}\left[ \tilde{\Gamma}_i \rho - \tilde{\gamma}_i \right] \right| \leq \epsilon'$.\\
\textbf{Theorem: } Let $\rho \in \left\{ \rho \in \mathcal{D}(\mathcal{H}_A \otimes \mathcal{H}_B^{N_c}) ~ : ~ \left| \text{Tr}\left[ \tilde{\Gamma}_i \rho - \tilde{\gamma}_i \right] \right| \leq \epsilon' \right\}$ where $\epsilon' > 0$ and $0 < \epsilon \leq \frac{1}{e(\text{dim}(\mathcal{G}(\rho))-1)}$. Then
\begin{equation}
    \min_{\rho \in \mathcal{S}}f(\rho) \geq \beta_{\epsilon \epsilon'}(\rho) - \zeta_{\epsilon}
\end{equation}
where $\zeta_{\epsilon} := 2\epsilon(\text{dim}(\mathcal{G}(\rho))-1) \log\left( \frac{\text{dim}(\mathcal{G}(\rho))}{\epsilon(\text{dim}(\mathcal{G}(\rho))-1)} \right)$ and 
\begin{equation}
\begin{aligned}
    \beta_{\epsilon, \epsilon'}(\sigma) :=& f_{\epsilon}(\sigma) - \text{Tr}\left[ \sigma^{\top} \nabla f_{\epsilon}(\sigma) \right] \\
    &+ \max{(\vec{y},\vec{z} \in \tilde{\mathcal{S}}_{\epsilon}^*(\rho)} \left( \vec{\tilde{\gamma}} \cdot \vec{y} - \epsilon' \sum_{i=1}^{|I|} \right).
\end{aligned}
\end{equation}
The set $\tilde{\mathcal{S}}_{\epsilon}^*(\sigma)$ is given by
\begin{equation}
\begin{aligned}
    &\tilde{\mathcal{S}}_{\epsilon}^*(\rho) :=\\
    &\left\{ (\vec{y}, \vec{z}) \in (\mathbb{R}^{|I|}, \mathbb{R}^{|I|}) ~|~ -\vec{z}  \leq \vec{y} \leq \vec{z}, ~ \sum_{i=1}^{|I|} y_i \tilde{\Gamma}_i^{\top} \leq \nabla f_{\epsilon}(\sigma) \right\}.
\end{aligned}
\end{equation}
We use this theorem to obtain a reliable lower bound.
%--------------------------------------------------------
%             Remarks on the implementation
%--------------------------------------------------------

\section{Remarks on the implementation}\label{sec:remarks}
It remains to specify expressions and protocol- or implementation-specific details that were treated in a general way, when summarising the idea of the security proof in Section \ref{sec:proof_method}.\\
First, we specify Bob's measurement operators $R_z$ (see eq. (\ref{eq:RegionOpDef}) ). Inserting the sets describing Bob's key map, given in eq. (\ref{eq:regions}). Bob performs heterodyne measurement, hence, according to \cite{Sanders_2004}, the POVM reads $E_y = \frac{1}{\pi} |\gamma\rangle \langle \gamma|$, where $\gamma \in \mathbb{C}$. Then, we obtain 
\begin{equation}\label{eq:Rra}
    R_z := \int_{A^{\text{r}}_z} E_{\gamma} \,d^2\gamma= \frac{1}{\pi} \int_{A^{\text{r}}_k} |\gamma\rangle\langle \gamma| \,d^2\gamma.
\end{equation}
Next, we express the region operators in the Fock basis 
\begin{align}\label{eq:R_ra_inf}
R_z &= \sum_{n=0}^{\infty} \sum_{m=0}^{\infty} \langle n |R_z |m\rangle |n\rangle\langle m|,
\end{align}
where we replace the upper limits by $N_c$ due to the photon-number cutoff assumption \cite{Lin_2019}. In \cite{Kanitschar_2021}, we derived analytical expressions for the region operators for different QPSK protocols. By a similar calculation with adapted angular integration, we obtain for the present 8PSK protocol
\begin{align}
&\langle n |R_z |m\rangle = \frac{1}{\pi} e^{-i(m-n)z\frac{\pi}{4}}\\ & \cdot \left\{
\begin{array}{ll}
\frac{\Gamma\left(n+1, \Delta_r^2 \right) }{ n!} \left(\frac{\pi}{8} - \Delta_a\right) & n = m \\
\frac{\Gamma\left(\frac{m+n}{2}+1, \Delta_r^2 \right) }{(m-n)\sqrt{n!}\sqrt{m!}}  \sin\left[\left(\frac{\pi}{8}-\Delta_a \right)(m-n) \right] & \, n \neq m \\
\end{array}
\right. .
\end{align}
Note that $\Delta_a$ is an angular postselection parameter, since in \cite{Kanitschar_2021} we investigate also angular postselection. In the present paper, we chose to investigate only radial postselection further, i.e., we perform no optimisation over $\Delta_a$ and set $\Delta_a = 0$ to reduce computation-time, as the problem is already very high-dimensional.\\

We carried out the coding for the numerical security proof in MATLAB\texttrademark R2020a and used CVX \cite{cvx1, cvx2} to model the linear semi-definite programs that appear in step 1 and step 2. Furthermore, we used the SDPT3 \cite{SDPT3_a, SDPT3_b} solver and the MOSEK solver \cite{mosek} to dispense the (SDP) optimisation tasks. We applied the modified Frank-Wolfe algorithm given in \cite{Winick_2018} to solve the linearised optimisation problem, where we used the bisection method to solve the line-search at the end of every Frank-Wolfe iteration.\\ 
The initial value, required to start the Frank-Wolfe algorithm, was calculated using a model for a two-mode Gaussian channel \cite{Weedbrook_2012} with excess noise $\xi$ and transmittance $\eta = 10^{-0.02L}$. This corresponds to a transmittance of $-0.2$dB or about $95.5\%$ per kilometer, which is a realistic value for practical implementations.\\
As mentioned in Section \ref{sec:protocol}, in the whole paper, we assume that Alice prepares her states according to the uniform distribution, hence $\forall l \in \{0,...,7\}:~ p_l = \frac{1}{8}$, which is expected to be the most efficient choice due to symmetry reasons. If not mentioned otherwise, we chose the cutoff number $N_c = 14$. It would have been possible to use a lower value, e.g., $N_c = 12$ for high transmission distances, where $|\alpha|$ is smaller than $1$, but for consistency reasons, we used the same cutoff number for all data points in the same curve. The maximal number of Frank-Wolfe steps varied between $N_{FW} = 10$ and $N_{FW} = 200$, depending on the chosen system parameters (mainly on the excess noise $\xi$ and the transmission distance $L$). The threshold for the Frank-Wolfe algorithm was chosen to be $\epsilon_{FW} = 10^{-7}$ and we chose the perturbation $\tilde{\epsilon}= 10^{-11}$.

%--------------------------------------------------------
%             Implementation and Validation
%--------------------------------------------------------

\section{\label{sec:Validation}Validation of the method}
Before we come to the results of our investigation, we validate our implementation. In the whole section, we used a reconciliation efficiency of $\beta = 0.95$ and set the excess-noise to $\xi = 10^{-5}$ for the numerical calculations (instead of $\xi = 0$) to improve the numerical stability. Furthermore, in the whole section, we do not perform any postselection, so $\Delta_r = 0$. Recall, that we chose the photon cutoff number $N_c = 14$. \\
We compare our data to the secure key rates, obtained by analytical calculation, generalising the analytical approach from \cite{Heid_2006} to eight-state protocols. That security proof considers the generalised beam-splitter attack, which is known to be optimal for loss-only (i.e., without noise) channels. Therefore, when Alice prepares a coherent state $|\alpha\rangle$, Bob receives another coherent state  $|\sqrt{\eta}\alpha\rangle$, where $\eta$ is the transmittance of the channel, while Eve receives the coherent-state $|\sqrt{1-\eta} \alpha\rangle$.\\

\begin{figure}[th!]
\subfloat[ \label{fig:alpha_th_vs_num_ideal}Optimal analytical $|\alpha|$ for different transmission lengths $L$ and for $\xi = 0$, found by fine-grained search in steps of $\Delta_{|\alpha|} = 0.005$ (red line) and optimal coherent state amplitudes obtained by numerical calculations and fine-grained search in steps of $\Delta_{|\alpha|} = 0.02$ (blue dots).]{
\includegraphics[width=0.5\textwidth]{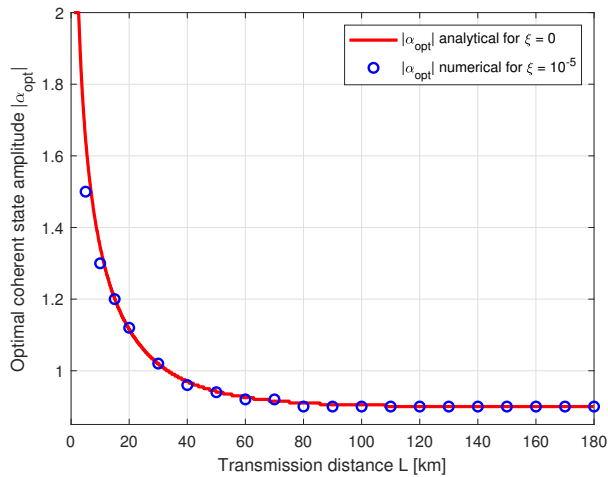}}\\
\subfloat[\label{fig:th_calc}We plot the analytical prediction (red dot-dashed line) and the second steps from our numerical calculations for transmission distances between $5$ and $180$km. The primary (left) y-axis displays the obtained secure key rate, while the secondary (right) y-axis displays the relative difference between the theoretical calculation and the numerical results (purple pluses). ]{
\includegraphics[width=0.5\textwidth]{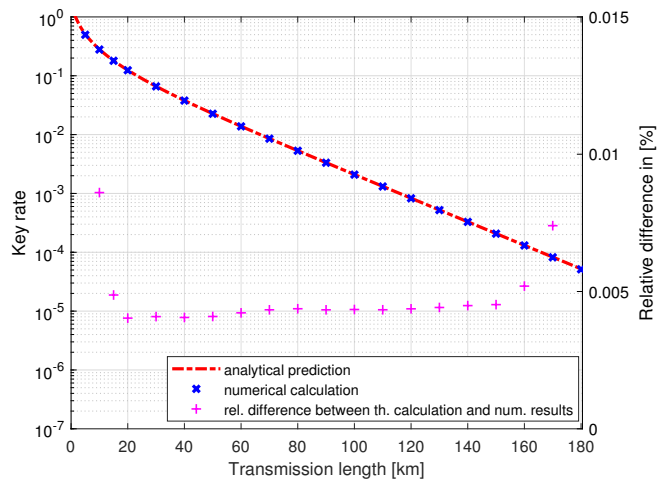}}
\caption{Comparison between key rates from theoretical prediction and the lower bounds on the secure key rate from our numerical results (almost) without noise and without performing postselection. The investigation shows that both the optimal coherent state amplitudes and the secure key rate obtained by numerical calculation match perfectly with the analytical prediction.\label{fig:Validation}}
\end{figure} 

First, we have to find the optimal choice for the coherent state amplitude $|\alpha|$, which might differ from theoretical prediction for the loss-only channel (see Figure \ref{fig:alpha_th_vs_num_ideal}). We carry out a fine-grained search in steps of size $0.02$ in the interval $|\alpha| \in [0.70, ´1.70]$, where the interval is chosen slightly bigger than the relevant range for $|\alpha|$ according to the theoretical prediction, to take possible changes in the optimal coherent state amplitude due to the non-zero excess-noise into account. We observe an excellent accordance between the theoretical prediction and our numerical results for the optimal coherent state amplitude $|\alpha|$ for all examined transmission distances, as can be seen in Figure \ref{fig:alpha_th_vs_num_ideal}.\\

 In Figure \ref{fig:th_calc}, we compare the secret key rates, obtained by the present numerical method with the analytical prediction for the noiseless channel for transmission distances between $5$km and $180$km. Again, one observes an excellent accordance with the analytical prediction for all distances. While the relative difference (purple pluses in Figure \ref{fig:th_calc}) for $5$km and $180$km exceed $1\%$, the relative differences for transmission distances between those values are lower than $0.01\%$, underlining the excellent accordance with the theoretical prediction. Furthermore, we note that the difference between the first- and the second step (so, between the upper- and lower bound on the secure key rate) is two magnitudes smaller than the secure key rate, indicating a tight gap between the upper and lower bound on the secure key rates. We note that the gap between the first and the second step increases for very low and very high transmission distances. This explains the relative differences greater than $1\%$ between our result and the theoretical prediction for $5$km and $180$km. Summing up, for the noiseless channel, our results are very satisfying, indicating a high reliability of our implementation.

%--------------------------------------------------------
%                       Results
%--------------------------------------------------------
\section{Results}\label{sec:results}

\begin{figure}
\centering
\includegraphics[width=0.48\textwidth]{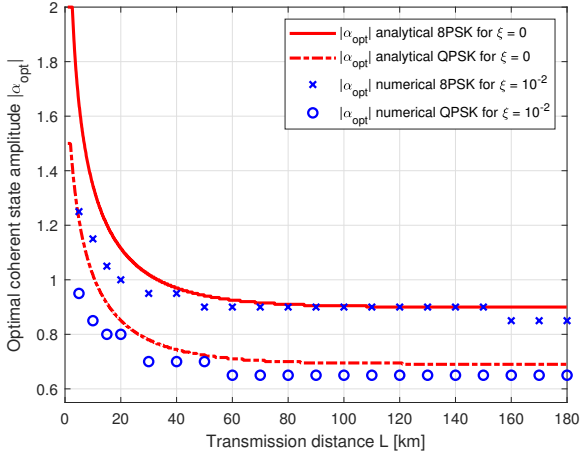}\caption{\label{fig:8PSK_alpha_xi_001} Optimal choice of the coherent state amplitude $|\alpha|$ for $\xi > 0$ obtained by coarse-grained search compared to predicted optimal choice for loss-only channel for QPSK and 8PSK protocol. As our results for $\xi = 0.01$ and $\xi > 0.01$ do not differ significantly, we plot only the data points for $\xi = 0.01$ to improve clarity. We are going to use those optimal values for $|\alpha|$ in what follows.}
\end{figure}

Again, the first task is to find the optimal choice for the coherent state amplitude $|\alpha|$ for channels with noise $\xi > 0$, which might differ from the theoretical prediction for the loss-only channel (see Figure \ref{fig:alpha_th_vs_num_ideal}). Therefore, we carry out a course-grained search in steps of size $0.05$ in the interval $|\alpha| \in [0.75, 1.70]$, where the interval is chosen slightly bigger than the relevant range for $|\alpha|$ according to the theoretical prediction. We investigate lengths between $5$km and $250$km and excess-noise levels of $\xi = 0.01$ and $\xi = 0.02$, while we do not perform any postselection. In  Figure \ref{fig:8PSK_alpha_xi_001}, we plot the results of the coarse-grained search and give the theoretical prediction (see Section \ref{sec:Validation}) as reference-curve. Since both the values of the theoretical prediction and the results of our numerical investigation remain constant for transmission distances higher than $80$km and $110$km respectively, we omit the part of the plot, exceeding $180$km. As expected, the optimal coherent state amplitude for noisy channels is slightly lower than those for a loss-only channel, but is still close to the theoretical prediction. We observe that the optimal choice for the coherent state amplitude decreases with increasing transmission distance, hence with increasing losses. This is in accordance with our expectations, as for high losses Eve receives a much stronger signal than Bob, whose signal has to pass the whole optical fibre, while Eve is assumed to extract Alice's signal right after leaving her lab. Hence, the amplitude has to be small for high transmission distances to keep Eve's advantage as small as possible. Based on our observations, it turns out that it is sufficient to search $|\alpha_{\text{opt}}|$ around the theoretical prediction. Furthermore, we note that the found optimal $|\alpha|$ does not differ significantly for all examined values of excess noise. We will use these optimal values for the coherent state amplitude for all calculations in the present chapter, if not stated otherwise.

\subsection{Secret key rates for 8PSK protocols}
Next, we examine the achievable secret key rates, using the optimal values for the coherent state amplitude from the previous section. We examine two different values of excess noise and two different reconciliation efficiencies.

\subsubsection{Comparison with secure key rates obtained with QPSK protocol}
\begin{figure}
\subfloat[$\xi = 0.01$, $\beta = 0.90$\label{fig:comp_w_QPSK_xi_001_beta_90}]{
    \includegraphics[width=0.48\textwidth]{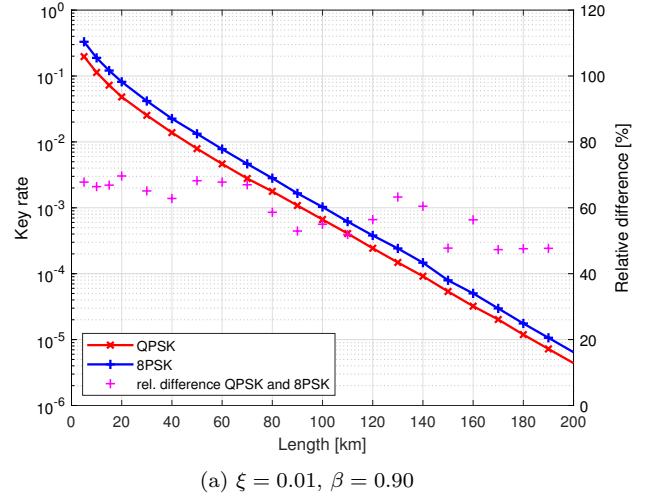}}\\
\subfloat[$\xi = 0.01$, $\beta = 0.95$\label{fig:comp_w_QPSK_xi_001_beta_95}]{
    \includegraphics[width=0.48\textwidth]{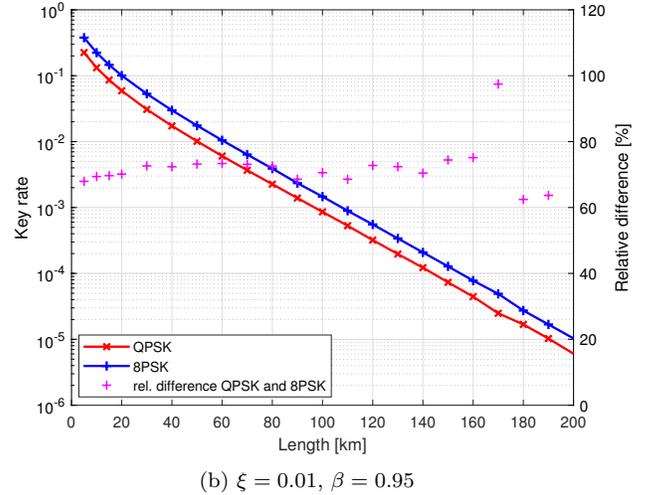}}
\caption{Comparison of the obtained secure key rates for QPSK and 8PSK protocol without performing postselection for two different values of $\xi = 0.01$ and $\beta \in \{0.90, 0.95 \}$. The secondary y-axis shows the relative difference between the lower bounds on the secure key rate, calculated for the 8PSK and the QPSK protocol.  \label{fig:comp_w_QPSK_xi_001} }
\end{figure}

\begin{figure}
\subfloat[$\xi = 0.02$, $\beta = 0.90$\label{fig:comp_w_QPSK_xi_002_beta_90}]{
    \includegraphics[width=0.48\textwidth]{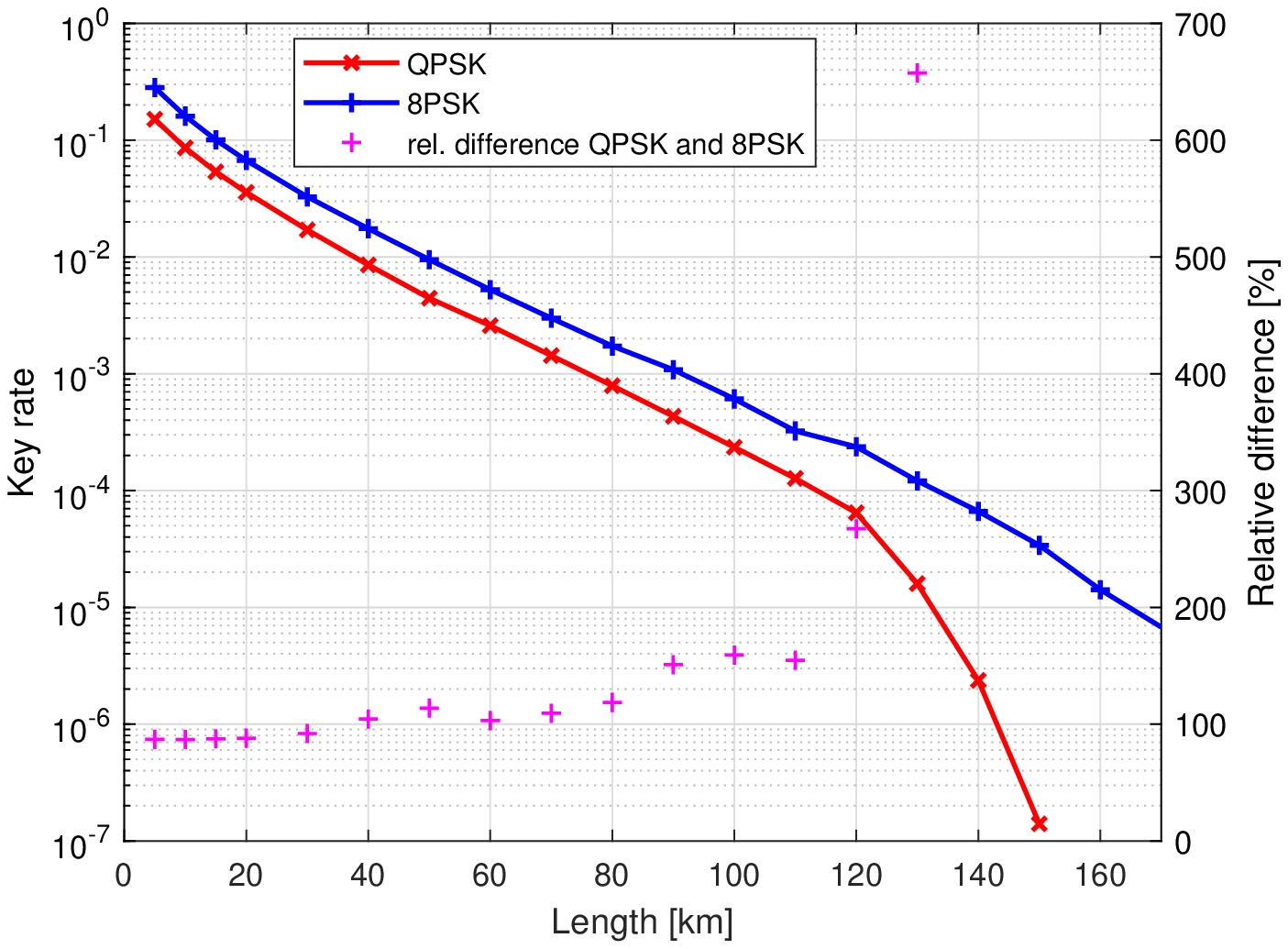}}\\
\subfloat[$\xi = 0.02$, $\beta = 0.95$\label{fig:comp_w_QPSK_xi_002_beta_95}]{
    \includegraphics[width=0.48\textwidth]{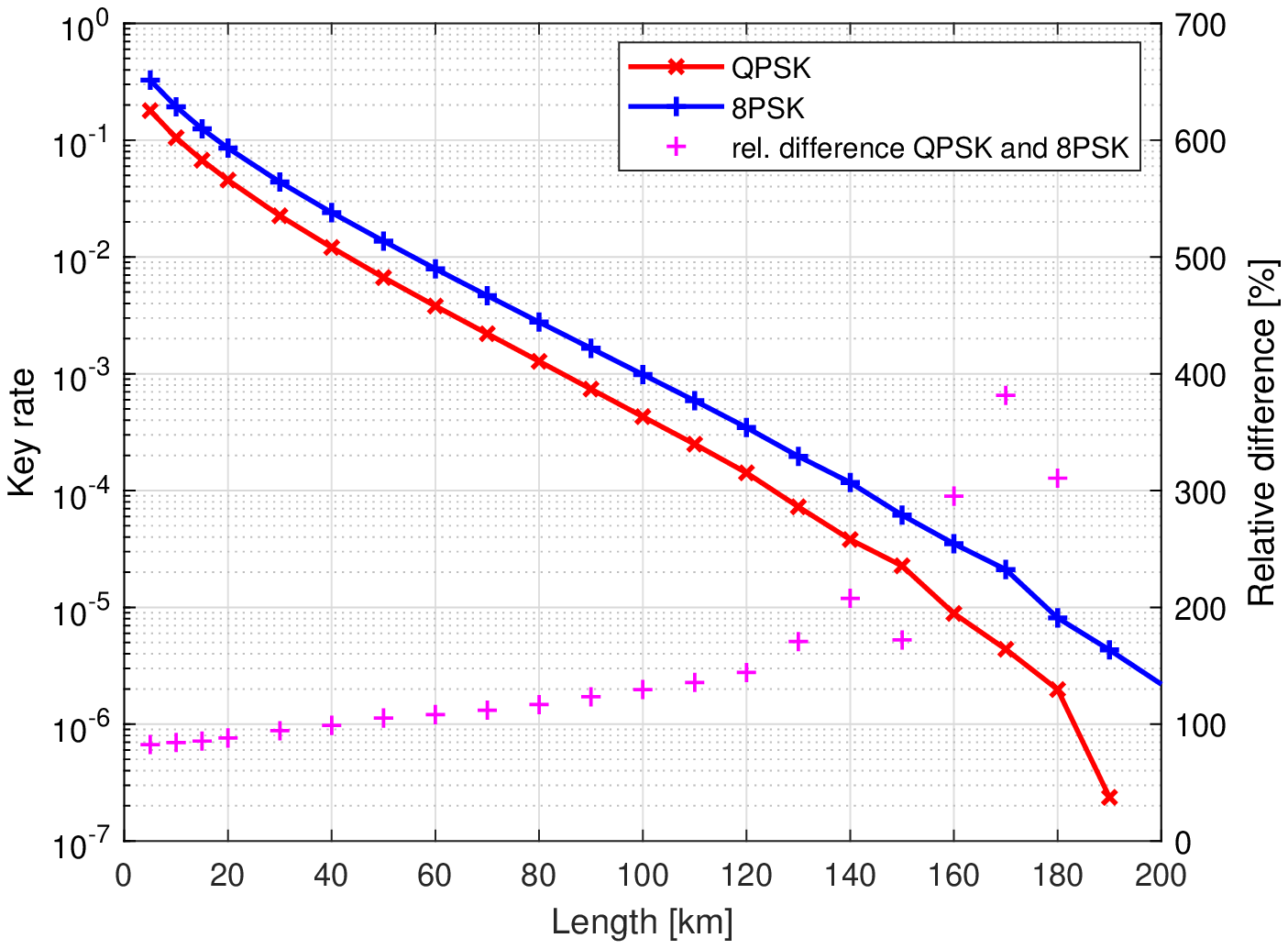}}
\caption{Comparison of the obtained secure key rates for QPSK and 8PSK protocol without performing postselection for two different values of $\xi = 0.02$ and $\beta \in \{0.90, 0.95 \}$. The secondary y-axis shows the relative difference between the lower bounds on the secure key rate, calculated for the 8PSK and the QPSK protocol.  \label{fig:comp_w_QPSK_xi_002} }
\end{figure}

First, we give the obtained secret key rates for transmission lengths up to $200$km and various values of excess-noise without any postselection and compare the results to the achieved key rates for the QPSK protocol without postselection. In Figures \ref{fig:comp_w_QPSK_xi_001} and \ref{fig:comp_w_QPSK_xi_002}, we show the obtained secure key rates for $\xi = 0.01$ and $\xi = 0.02$ and reconciliation efficiencies of $\beta = 0.90$ and $\beta = 0.95$ for both protocols and plot the relative difference between the key rates on the secondary y-axis.\\
For $\xi = 0.01$, one observes an outperformance of the 8PSK protocol compared to the QPSK protocol of about $45$ to $68\%$ if the reconciliation efficiency is chosen to be $\beta = 0.90$ and of about $60$ to $80\%$ for $\beta = 0.95$, where the relative difference shows only a little dependency on the transmission distance.\\
For $\xi = 0.02$, we obtained higher relative differences of about $100\%$ for transmission distances up to $110$km for both values of reconciliation efficiency $\beta$. For higher transmission lengths, the relative difference increases notably as the higher noise level causes a significant drop in the secure key rate for the QPSK protocol. We note that for transmission distances of $140$km and higher the gap between step 1 and step 2 increases slightly for the eight-state protocol, while the gap remains small for the four-state protocol, except for the regions where the key rates drop steeply. This is due to numerics and may be improved. As step 2 serves as a lower bound on the key rate, this decreases the bound for the 8PSK protocol. Therefore, we expect even a slightly higher outperformance (i.e., a higher relative difference) for transmission distances greater than $140$km if the gap can be narrowed down. Missing relative differences for data points where the QPSK key rates drop steeply indicate that the corresponding points exceed the scale in the plots, which was chosen such that the majority of the points can be read well.\\


\subsubsection{Examining the influence of radial postselection}
\begin{figure}
\subfloat[$\xi = 0.01$, $\beta = 0.90$\label{fig:comp_noPS_rPS_xi_001_beta_90}]{
    \includegraphics[width=0.48\textwidth]{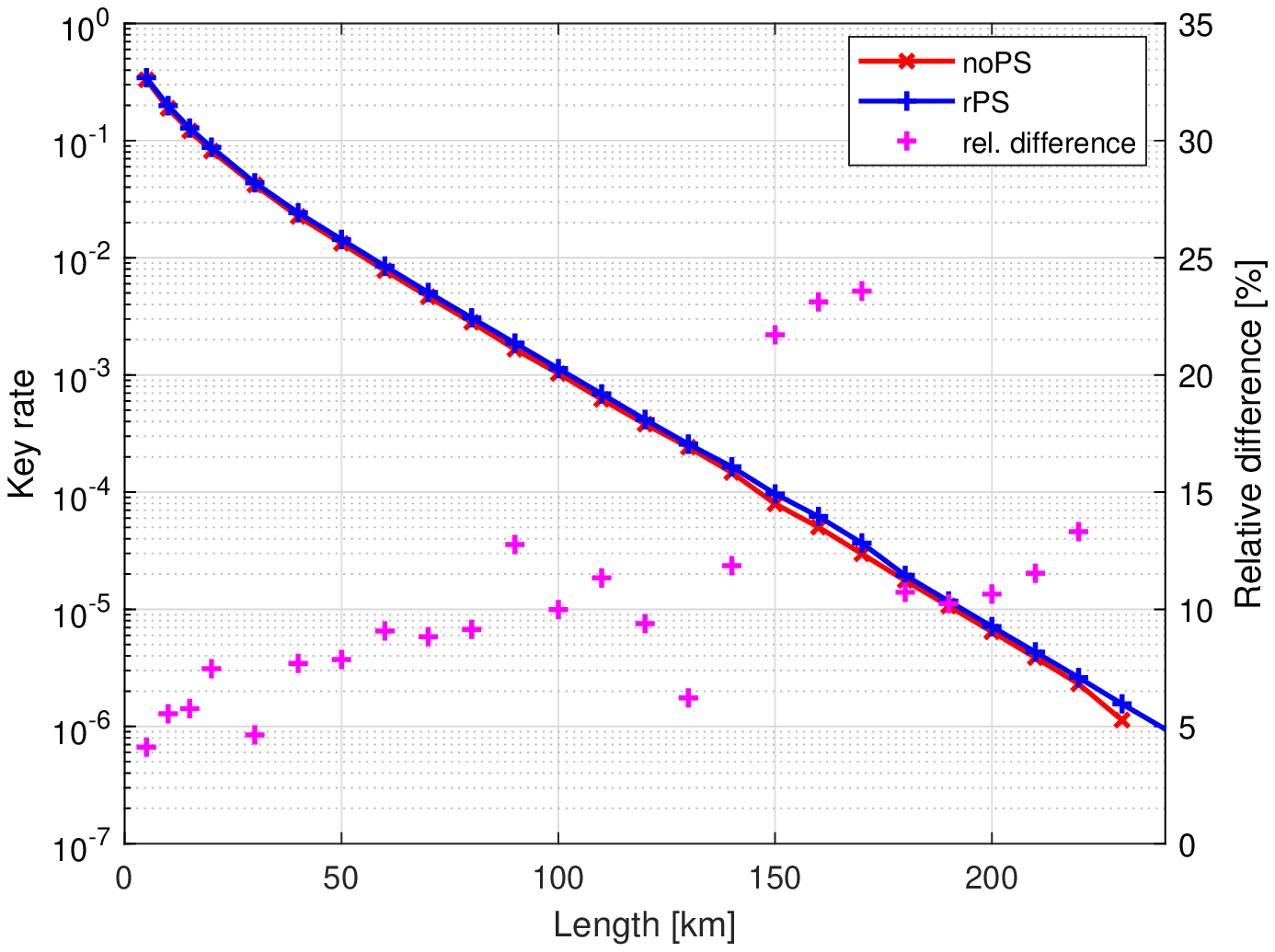}}\\
\subfloat[$\xi = 0.01$, $\beta = 0.95$\label{fig:comp_noPS_rPS_xi_001_beta_95}]{
    \includegraphics[width=0.48\textwidth]{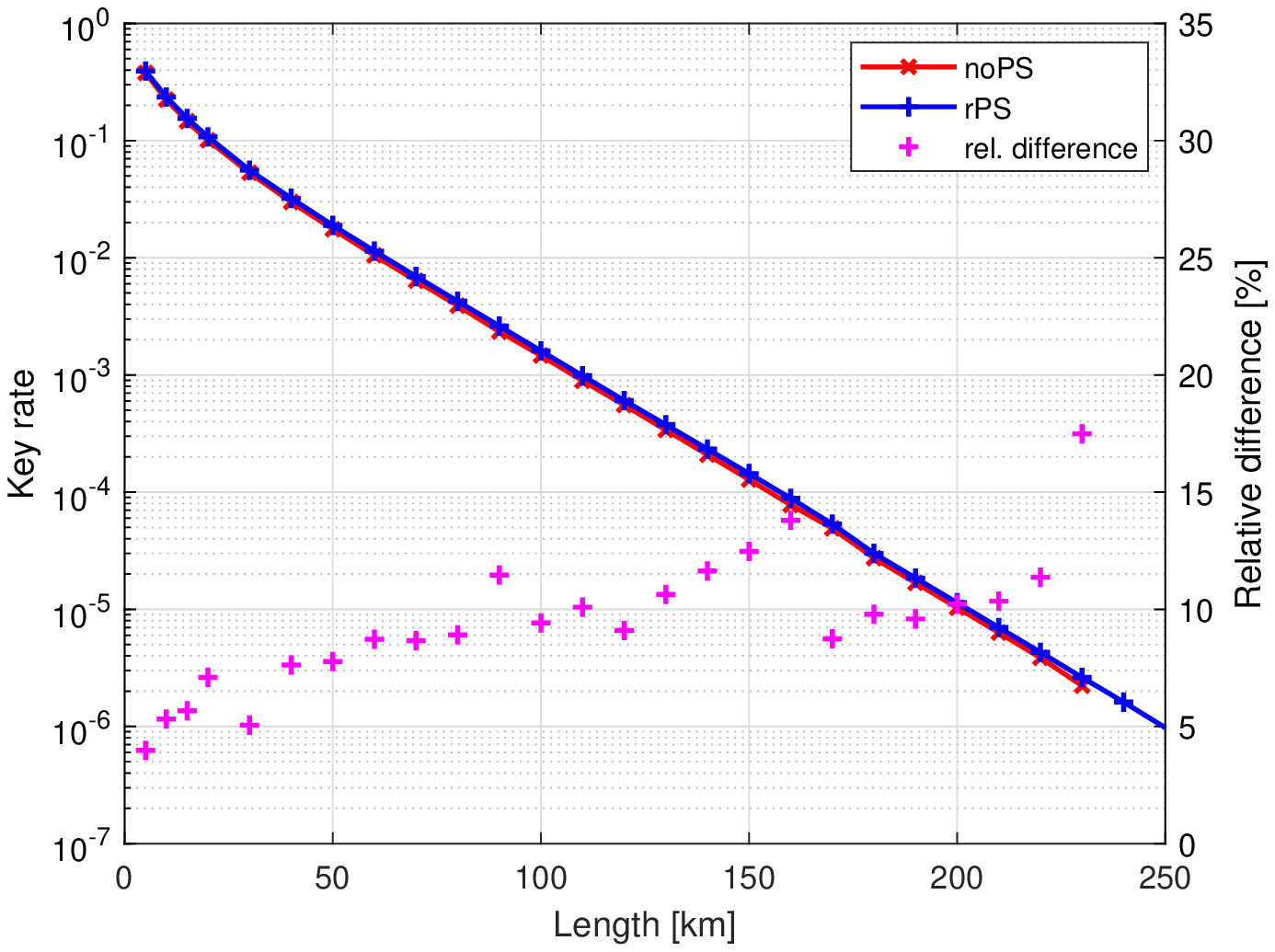}}
\caption{\label{fig:comp_noPS_rPS_xi_001} Comparison of secure key rates for transmission distances up to $250$km between an 8PSK protocol without postselection and with radial postselection for $\xi = 0.01$ and $\beta \in \{0.90, 0.95 \}$. The secondary y-axis displays the relative difference between the key rates obtained with radial postselection and without postselection. Missing data points correspond to data point where the calculation for the protocol without any postselection did not lead to positive key rates after the second step. }
\end{figure}

\begin{figure}
\subfloat[$\xi = 0.02$, $\beta = 0.90$\label{fig:comp_noPS_rPS_xi_002_beta_90}]{
    \includegraphics[width=0.48\textwidth]{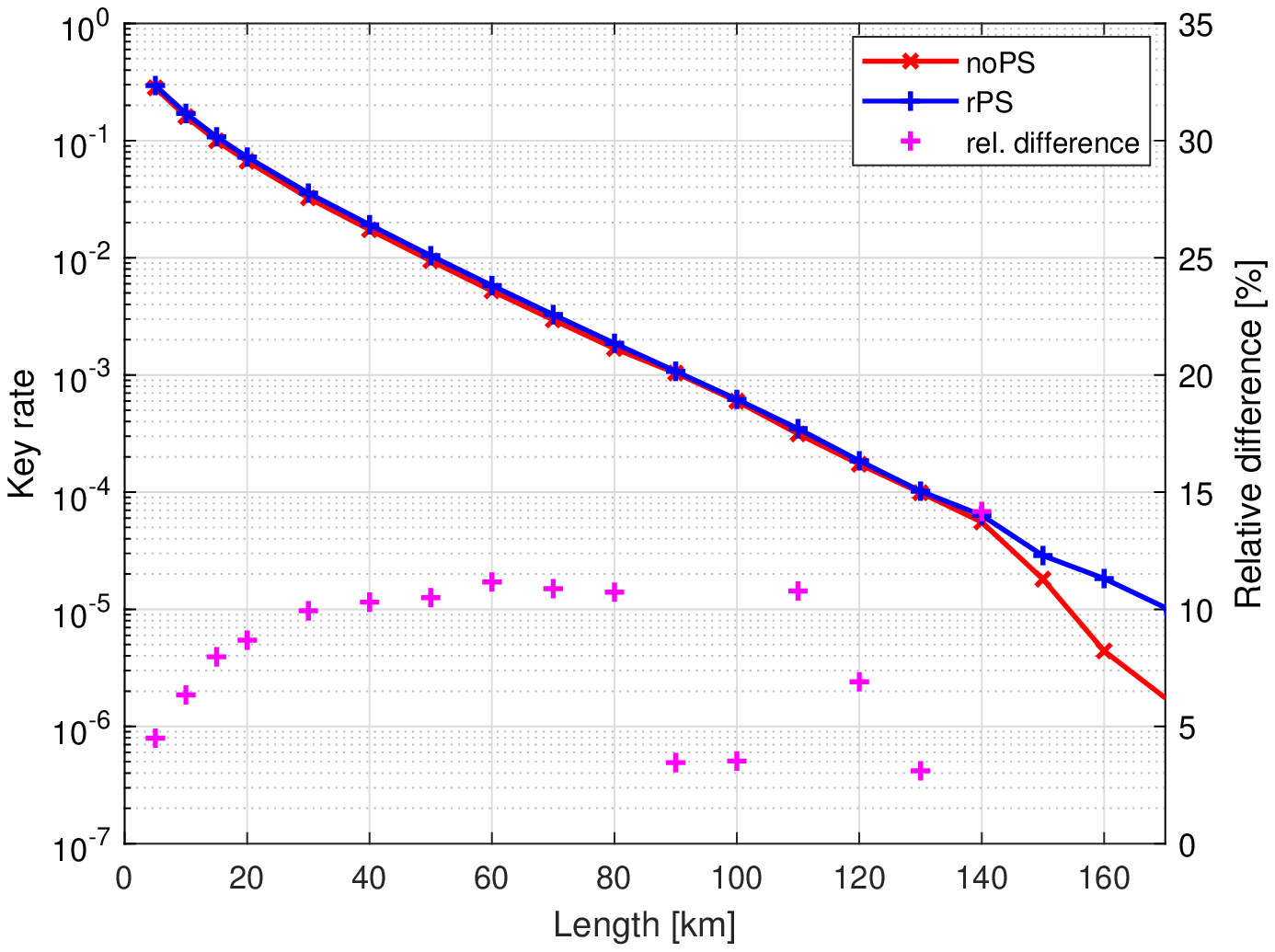}}\\
\subfloat[$\xi = 0.02$, $\beta = 0.95$\label{fig:comp_noPS_rPS_xi_002_beta_95}]{
    \includegraphics[width=0.48\textwidth]{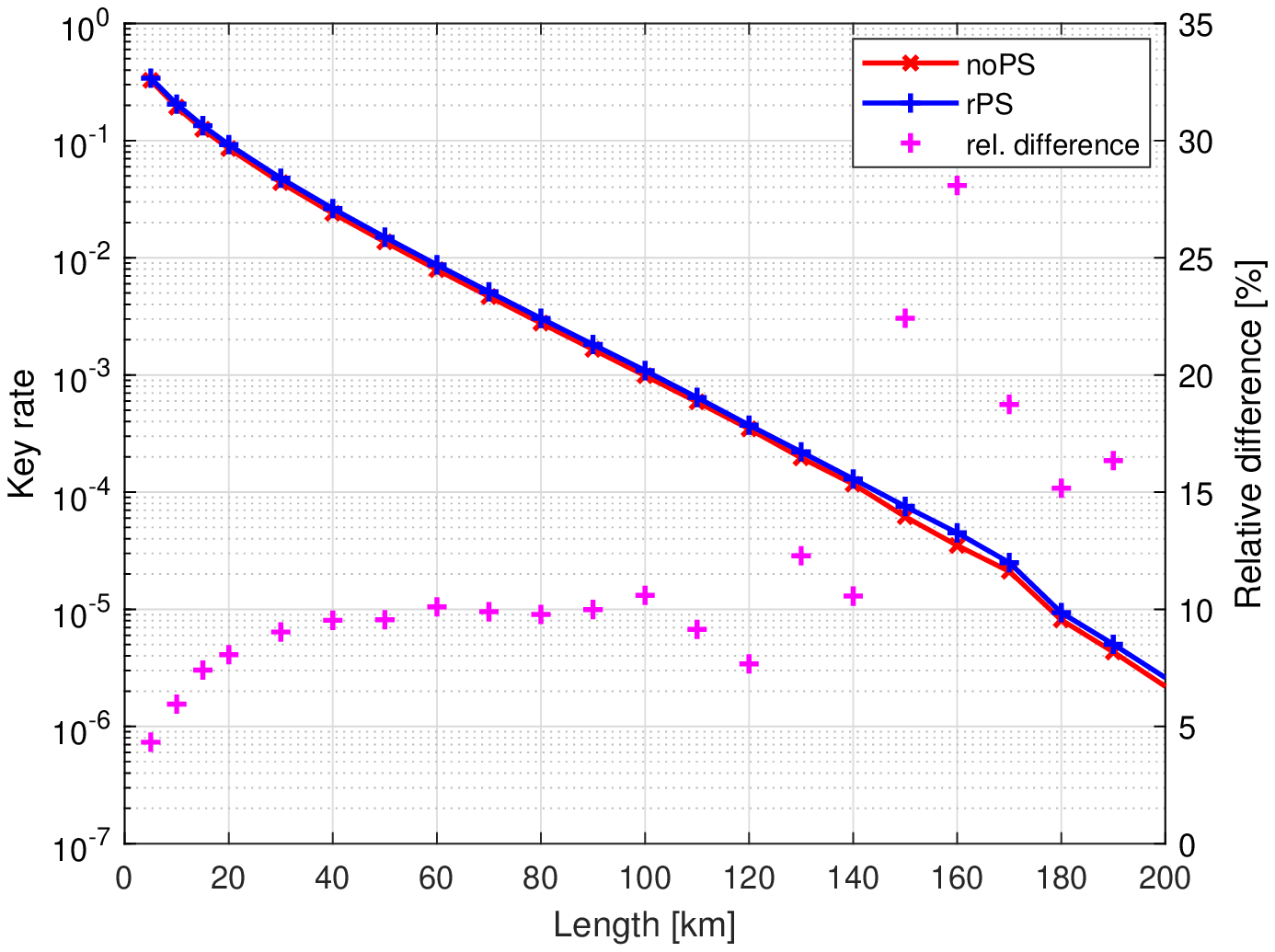}}
\caption{\label{fig:comp_noPS_rPS_xi_002} Comparison of secure key rates for transmission distances up to $200$km between an 8PSK protocol without postselection and with radial postselection for $\xi = 0.02$ and $\beta \in \{0.90, 0.95 \}$. The secondary y-axis displays the relative difference between the key rates obtained with radial postselection and without postselection. Missing data points correspond to data point where the calculation for the protocol without any postselection did not lead to positive key rates after the second step.  }
\end{figure}

Second, we examine the influence of postselection on the secure key rate. By construction, postselection omits some fraction of the measurement results. It is expected to contribute significantly to the overall key rate, in particular for high transmission distances, where the signal losses are high, hence Eve has a much stronger signal than Bob. This is because Eve is assumed to grab her share of the signal before it enters the quantum channel, hence her signal has not experienced any losses or noise, while Bob's signal did. Therefore, the communicating parties try to reduce Eve's information about the key by lowering the signal amplitude and using postselection to reduce parts of the key, where Eve might have gained more information than the communicating parties, by postselection. Additionally, we expect postselection also to be advantageous in order to mitigate Eve's edge due to the channel noise. While postselection is expected to have a positive influence on the key rate, obviously, choosing the postselection-areas too large, results in a decrease in the secure key rate, as parts of the key that could have been used to generate a secret key were omitted. Therefore, one can expect to find some sweet-spot, where the secure key rate can be increased maximally. Note that the case without postselection is included in every postselection scheme by setting the postselection parameter $\Delta = 0$. As the calculations for eight-state protocols involve solving very high-dimensional semi-definite programs, hence are computationally very expensive, we chose to focus on the investigation of influence of the radial postselection parameter $\Delta_r$ on the secret key rate.Therefore, we perform coarse-grained search and vary $\Delta_r$ with step-sizes of $0.05$ in the interval $\Delta_r \in [0, 0.65]$ which turned out to be sufficient to find the maximal key rate.\\
In Figure \ref{fig:comp_noPS_rPS_xi_001}, we plotted both the result without postselection and with radial postselection for transmission distances up to $250$km (in steps of $10$km) for excess-noise levels of $\xi = 0.01$ and $\xi = 0.02$. The secondary y-axis shows the relative difference between both curves. For $\xi=0.01$ and $\beta = 0.95$ we observe relative differences of about $5\%$ for very low transmission distances and up to $14\%$ for high transmission lengths, while the relative outperformance starts at $5\%$ for low transmission distances and goes up to $25\%$ for medium to high transmission distances for $\beta = 0.90$. We see qualitatively similar results for $\xi = 0.02$ in Figure \ref{fig:comp_noPS_rPS_xi_002}. As expected, based on our findings for four-state protocols, radial postselection increases the secure key rate for all transmission distances, as the case without postselection is included in the radial postselection scheme by setting $\Delta_r = 0$. Furthermore, we observe an increasing impact on the secure key rate for high transmission distances, and higher values of excess noise. We note that the exact value of the relative outperformance is influenced by the gap between step 1 and step 2, in particular for higher transmission distances. Therefore, we expect that numerical improvements would lead to smoother absolute- and relative difference curves.


\subsection{Influence of the probability to pass the postselection}
In the previous section, we showed that one can increase the secure key rate of eight-state phase-shift keying protocols by applying radial postselection, and we investigated the magnitude of the increase in key rate. Besides maximising the secure key rate, experimentalists might aim to reduce the raw key to reduce the effort of the error-correction phase, which is known to be computationally expensive. Therefore, it can be interesting to examine the relation between the achievable secure key rate and the probability to pass the postselection phase $p_{\text{pass}}$ (or, alternatively, the probability of being postselected $1-p_{\text{pass}}$) in order to know either the reduction in raw key for some fixed (for example, the maximal) key rate or to know the key rate for some given raw key reduction. \\

\begin{figure}
\subfloat[$\beta = 0.90$\label{fig:p_pass_examination_90}]{
    \includegraphics[width=0.48\textwidth]{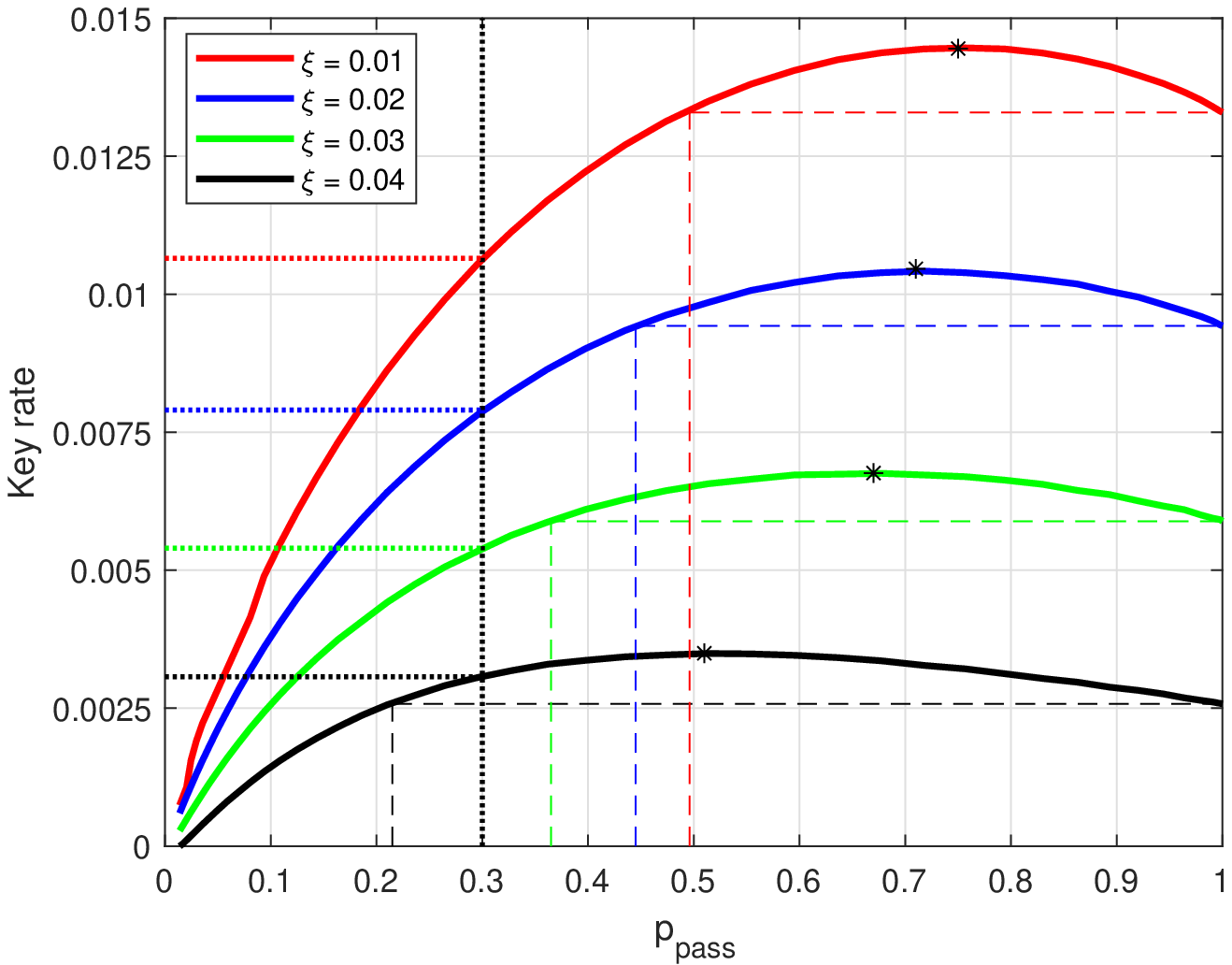}}\\
\subfloat[$\beta = 0.95$\label{fig:p_pass_examination_95}]{
    \includegraphics[width=0.48\textwidth]{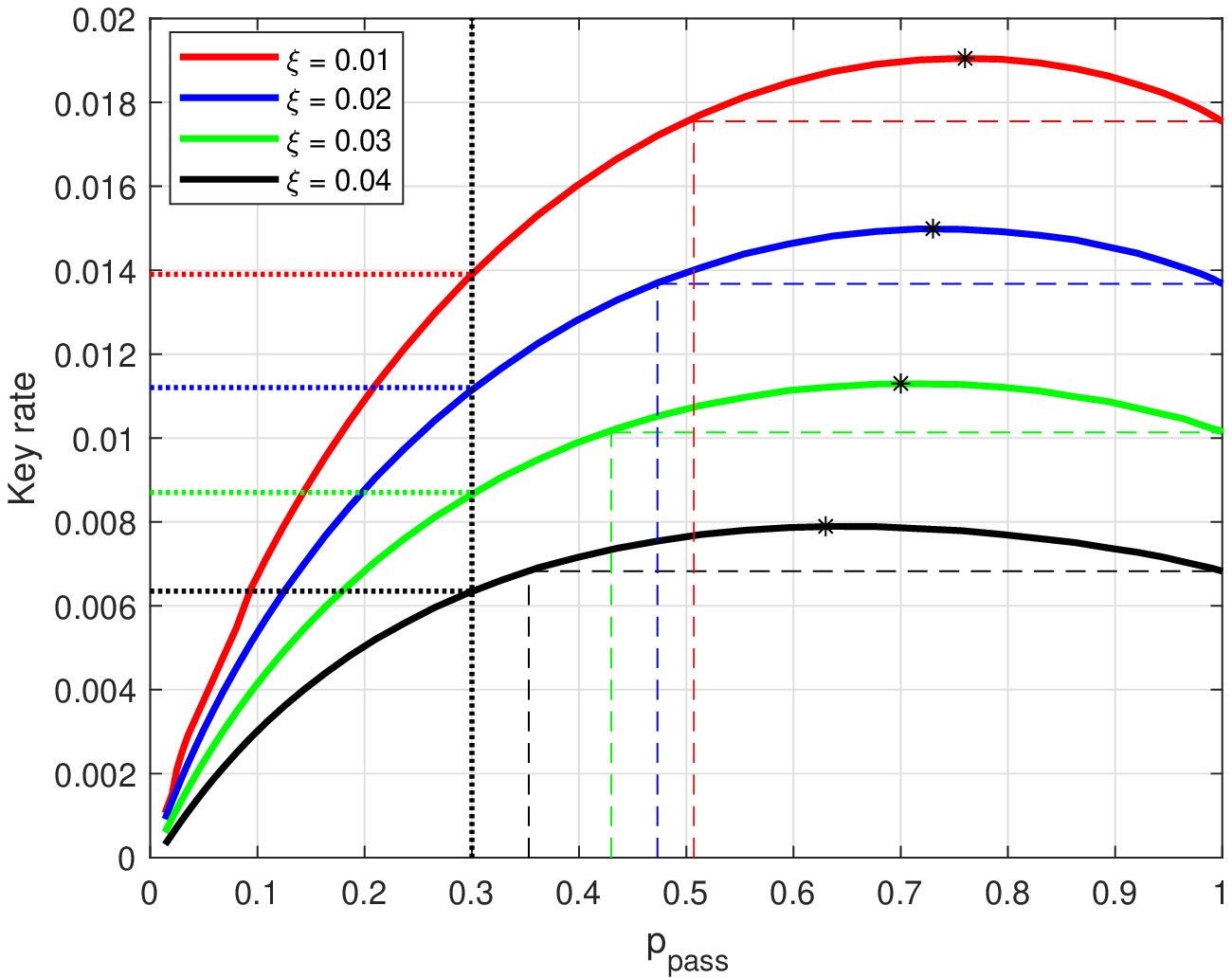}}
\caption{Secure key rate versus the probability to pass the postselection phase $p_{pass}$ for radial postselection and four different values of excess noise. The underlying data was calculated varying the postselection parameter $\Delta_r$ in the interval $[0,2.15]$ with a step size of $0.025$. \label{fig:p_pass_examination}}
\end{figure}
Therefore, we fixed $L=50$km and $|\alpha| = 0.90$ (which, according to Figure \ref{fig:8PSK_alpha_xi_001}, is the optimal value for $50$km) and varied the radial postselection parameter $\Delta_r$ in the interval $[0, 2.15]$ with a step-size of $0.05$. We investigated four different values of excess-noise, $\xi \in \{0.01, 0.02, 0.03, 0.04\}$, and two different values for the reconciliation efficiency, $\beta = 0.90$ and $\beta = 0.95$, which are the relevant values for many QKD systems. We note that $\beta = 0.95$ can be achieved with low-density parity-check codes. We plot our results in Figure \ref{fig:p_pass_examination}, where Figure \ref{fig:p_pass_examination_90} shows the secure key rates for $\beta = 0.90$ and Figure \ref{fig:p_pass_examination_95} shows the results for $\beta = 0.95$. As the gap between the first and the second step turned out to be very small, we plot only our results for the second step, as that serves as lower bound on the secure key rate. We note that the achieved maximal key rates for $\xi = 0.01$ and $\xi = 0.02$ coincide with those reported for $L=50$km in the previous section. \\

We proceed with a brief discussion of Figure \ref{fig:p_pass_examination}, where in Figure \ref{fig:p_pass_examination_90}, the reconciliation efficiency is $\beta = 0.90$ and in Figure \ref{fig:p_pass_examination_95} the reconciliation efficiency is $\beta =0.95$. We see that for all displayed curves the key rates attain their maximum for passing probabilities smaller than $1$. We observe that the maxima are shifted to the left for increasing excess-noise $\xi$, meaning that a higher noise-level requires more postselection to obtain the maximal key rate, which meets with our expectations. The curves motivate, three different strategies to reduce the raw key rate. First, one can aim to maximise the secure key rate, as discussed earlier. Then, the raw key is reduced moderately, while the key rate grows. Second, one can decide not to change the achievable secure key rate (as obtained without performing postselection) while reducing the raw key rate. This can be visualised by the intersection points of a horizontal line trough the secure key rate at $p_{\text{pass}} = 1$ with the key rate curve, as can be seen in Figures \ref{fig:p_pass_examination_90} and \ref{fig:p_pass_examination_95} (dashed lines). Third, one can decide to omit a certain fraction of the raw key, e.g., $70\%$, probably on the cost of a moderate decrease in the secure key rate. This scenario is visualised in Figures \ref{fig:p_pass_examination_90} and \ref{fig:p_pass_examination_95} with dotted lines. We list the passing probabilities corresponding to the first and second scenario in Table \ref{table:keyRate_vs_ppass} and the changes in the secure key rate for the third scenario with $p_{\text{pass}}$ fixed to $30\%$ in  Table \ref{table:keyRate__70_omit}.

\begin{table}[]
\begin{tabular}{|c||c|c||c|c|}
\hline 
      & \multicolumn{2}{c|}{$p_{\text{pass}}$ (at max. key rate)} & \multicolumn{2}{c|}{$p_{\text{pass}}$ (same key rate as noPS)} \\ \hline
$\xi$ & $\beta = 0.90$             & $\beta = 0.95$            & $\beta = 0.90$                 & $\beta = 0.95$                 \\ \hline \hline

0.01  & 0.75                       & 0.76                      & 0.50                           & 0.51                           \\ \hline
0.02  & 0.71                       & 0.73                      & 0.44                           & 0.47                           \\ \hline
0.03  & 0.67                       & 0.70                      & 0.36                           & 0.43                           \\ \hline
0.04  & 0.51                       & 0.63                      & 0.21                           & 0.34                           \\ \hline 

\end{tabular}\caption{Summary of results for secure key rate vs. probability to pass the postselection for two different values of $\beta$ and four different values of excess noise. The second and third column show the value of $p_{\text{pass}}$, when obtaining the maximal secure key rate. In the last two columns, one finds the value for $p_{\text{pass}}$ where the key rate has the same value as one obtains without performing postselection. So, one is left with exactly the share of the raw key given in the corresponding cell, while obtaining the same secure key rate as without performing postselection at all.}
\label{table:keyRate_vs_ppass}
\end{table}
\begin{table}[]
\begin{tabular}{|c||c|c|}
\hline
      & \multicolumn{2}{c|}{Change in secure key rate} \\ \hline
$\xi$ & $\beta = 0.90$         & $\beta = 0.95$        \\ \hline\hline

0.01  & $-20\%$                & $-21\%$               \\ \hline
0.02  & $-16\%$                & $-19\%$               \\ \hline
0.03  & $-8\%$                 & $-14\%$               \\ \hline
0.04  & $+19\%$                & $-8\%$                \\ \hline
\end{tabular}\caption{Change in the secure key rate when omitting $70$\% of the raw key compared to the secure key rate obtained without performing postselection for two different values of $\beta$ and four different values of excess noise.}
\label{table:keyRate__70_omit}
\end{table}

%--------------------------------------------------------
%                       Conclusions
%--------------------------------------------------------

\section{Conclusion and discussion \label{sec:Conclusion}}
We investigated the achievable secure key rates for the proposed 8PSK protocol with heterodyne measurement, using a recent numerical security proof technique, and showed that it yields about $70-80\%$ higher key rates than a QPSK protocol with a comparable protocol structure. Our results show that the maximal transmission distance can be improved for higher values of excess noise (here: $\xi = 0.02$), using the 8PSK protocol instead of the QPSK protocol.
Therefore, eight-state phase-shift keying protocols increase both the achievable secure key rate and the achievable range of continuous-variable quantum key distribution systems with phase-shift keying modulation. We showed that for the 8PSK protocol performing radial postselection can increase the secure key rate by up to approximately $14\%$ compared to no postselection. This value is similar to the effect of radial postselection in QPSK protocol \cite{Kanitschar_2021}. We showed that performing radial postselection in the 8PSK protocol that reduces the \emph{raw key rate} significantly by $50-80\%$ (depending on the level of excess-noise and the reconciliation efficiency) can result in the same \emph{secure key rate} as the 8PSK protocol without postselection. This addresses the high computational demand of the error-correction phase directly by reducing its input data, and can be implemented easily in software both in new and existing CV-QKD systems.

\begin{acknowledgements}
This work has received funding from the EU Horizon-2020 research and innovation programme under grant agreement No 857156 (OpenQKD) and 820466 (CiViQ).\\
\includegraphics[width=0.25\textwidth]{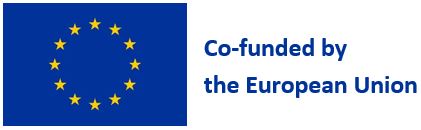} 

\end{acknowledgements}

%--------------------------------------------------------
%                       Appendix
%--------------------------------------------------------

\bibliography{Bibliography}

\end{document}